\documentclass[12pt,preprint]{aastex}

\shorttitle{Stellar Rotation in B-Stars}
\shortauthors{Huang \& Gies}

\begin{document}

\received{}
\accepted{}

\title{Stellar Rotation in Field and Cluster B-Stars}

\author{W. Huang}

\affil{Department of Astronomy \\
Caltech, MC 105-24, Pasadena, CA 91125;\\
wenjin@astro.caltech.edu}

\author{D. R. Gies}

\affil{Center for High Angular Resolution Astronomy\\
Department of Physics and Astronomy \\
Georgia State University, P. O. Box 4106, Atlanta, GA  30302-4106;\\
gies@chara.gsu.edu}


\slugcomment{Submitted to ApJ}

\paperid{73056}


\begin{abstract}
We present the results of a spectroscopic investigation of 108 nearby
field B-stars.  We derive their key stellar parameters,
$V \sin i$, $T_{\rm eff}$, $\log g$, and
$\log g_{\rm polar}$, using the same methods that we used
in our previous cluster B-star survey.  By comparing the results
of the field and the cluster samples, we find that the main
reason for the overall slower rotation of the field sample is that it
contains a larger fraction of older stars than found in the 
(mainly young) cluster sample.  
\end{abstract}

\keywords{line: profiles --- 
 stars: rotation ---
 stars: fundamental parameters ---
 stars: early-type
 }


\setcounter{footnote}{0}
\section{Introduction}                              
It is a curious but well known fact that field B-stars rotate 
slower than cluster B-stars \citep*{abt02,str05,hua06a,wol07},
but the explanation for the difference is still controversial.  
One possible solution is that field B-stars represent a population that
contains more evolved stars than cluster B-stars do. They appear to
rotate slower because stars generally spin down as they evolve
\citep{abt02,hua06a,hua06b}.  On the other hand, \citet{str05} and \citet{wol07}
suggest that difference in rotation rates between field and cluster
B-stars is mainly due to the difference between the initial conditions
of the stellar forming regions. The denser the environment (such as in
young open clusters), the more rapid rotators can form.  The second
explanation brings more attention to the possible connection between 
stellar rotation and the physical mechanisms playing a role during 
the star formation stage.  A plausible
higher accretion rate around a forming star in a denser region may
lead to a higher initial angular momentum and a shorter accretion
disk lifetime with its associated spin-down effects via magnetic
interactions between the star and the disk.

Because both the evolutionary status of stars and the initial conditions
of their forming regions may influence their rotation rates,
knowing the evolutionary status of these stars precisely becomes
a prerequisite for the solution of this puzzle.
With this in mind, we made a spectroscopic investigation
of 108 field B-stars using the same methods that we applied
in our previous cluster B-star survey \citep{hua06a,hua06b}.  There
are two advantages over previous studies of this topic: 1) Because 
we apply identical spectroscopic methods to both the field and 
cluster samples, the influence of any imperfection in our 
methods on the final comparisons will be reduced
to a minimum; 2) We use the estimated $\log g_{\rm polar}$ as 
an indicator of stellar evolutionary status, which is more
accurate and reliable for large numbers of stars with diverse
masses and rotation rates.  We describe our
derivation of the key stellar parameters of a field sample of B-stars
in next section.  The results of a comparison between the field
sample and the cluster sample are reported in Section 3, and a
short discussion and our conclusion are given in Section 4.


\section{Field B-Star Sample}                       

Our field B-star sample was selected from the NOAO Indo-U.S.\ Library
of Coud\'e Feed Stellar Spectra\footnote{http://www.noao.edu/cflib/}
\citep{val04}.  This library contains moderate resolution
spectra (FWHM = $1 - 2$\AA) of 1273 stars that were 
obtained with the 0.9-m Coud\'e Feed telescope at Kitt Peak National 
Observatory.  Roughly about 140 B-star spectra are found in this library.
These spectra are comparable in S/N and resolution to those analyzed 
in our previous cluster B-star survey.

Following the exact same procedure that we applied to cluster
B-stars \citep{hua06a,hua06b}, we obtained the stellar parameters
of 108 B stars in our final sample: 
the projected rotational velocity $V \sin i$, 
the effective temperature $T_{\rm eff}$, the apparent gravity $\log g$,
and the estimated polar gravity $\log g_{\rm polar}$.  
These results are summarized in Table 1.
We excluded all double-lined spectroscopic binaries (SB2) from the sample
because the derived parameters of these objects are not reliable.
The errors are estimated from the deviations between the observed and 
model profiles (see \citealt{hua06b}), and inclusion of uncertainties 
related to the continuum placement may increase these errors by $\approx 40\%$.

\placetable{tab1}      

The $V \sin i$ values were derived by fitting synthetic model profiles of \ion{He}{1}
$\lambda4471$ (or \ion{Mg}{2} $\lambda4481$ if the \ion{He}{1} line is too weak) to
the observed profiles, using realistic physical models of rotating stars
(including Roche geometry and gravity darkening).  The details of this
step are described in \citet{hua06a}.
One concern about the derived $V \sin i$ values is that we do not know the
exact instrumental broadening data of the NOAO Indo-U.S. Library 
spectra for the investigated region (4470 - 4480 \AA), and assumed only
the lower limit of the given FWHM range, 1 \AA, 
in the convolution of our synthesized line profiles.  
An underestimation of the instrumental broadening can lead to higher 
derived $V \sin i$ values.  In order to determine and then correct
the possible systematic errors caused by the uncertainty in 
the assumed instrumental broadening, we also obtained high resolution
spectra ($R = \lambda/\triangle\lambda \sim 42000$) 
of 34 stars in our sample from the ELODIE 
archive\footnote{http://atlas.obs-hp.fr/elodie/} \citep{mou04}.  
By comparing the $V \sin i$ values derived
from the NOAO library to those from the ELODIE library, we found the best
relationship between them can be written as 

\begin{equation}\label{eq_vsini_correct}
V \sin i_{\rm Elodie} \simeq \sqrt{(V \sin i_{\rm NOAO})^2 - (46 ~{\rm km~s}^{-1})^2}.
\end{equation}

For the stars in our sample that are not found in the Elodie library,
we corrected their $V \sin i$ using eq.~\ref{eq_vsini_correct} for $V \sin i_{\rm NOAO}
> 46$ km~s$^{-1}$, and we set $V \sin i = 0$ for $V \sin i_{\rm NOAO} \leq
46$ km~s$^{-1}$.  The corrected $V \sin i$ and its numerical fitting error
are given in columns (6) and (7)
of Table~1.  A comparison of the derived $V \sin i$ values between our results 
and those from \citet{abt02} is illustrated in Figure~1.  The good agreement in the
low $V \sin i$ region indicates that our corrections to $V \sin i$ are properly
assigned.  In the high $V \sin i$ region, our results are systematically greater
than the results from \citet{abt02}.  This discrepancy is not surprising, considering that our models
take the gravity darkening effect into account.  \citet*{tow04} showed
that the $V \sin i$ derived from fitting the \ion{He}{1} $\lambda4471$ line could
be lower by as much as 10-20\% for a rapid rotator if the strong gravity
darkening effect on its surface is ignored.  The most discrepant point
in Figure~1 is the star HD~172958.  Our $V \sin i$ measurement of this
star (167 km~s$^{-1}$) is similar to the measurements by \citet{pea87} and 
\citet{wol78} (175 km~s$^{-1}$).  The much larger value measured by \citet{abt02},
$V\sin i = 315$ km~s$^{-1}$, might result if the star is 
an unresolved, doubled-line binary that was observed
at a time of larger relative Doppler shifts, but the
star is not a known binary.

\begin{figure}
\epsscale{0.8}
\plotone{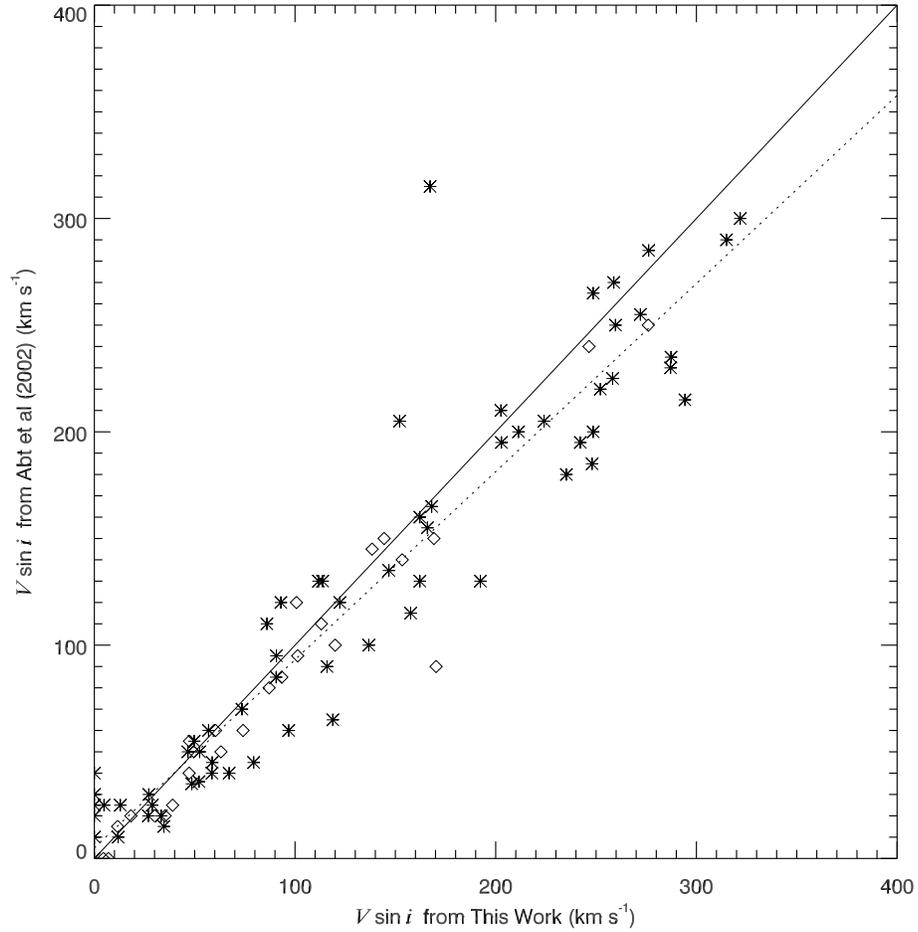}
\caption{
A comparison of our measured $V \sin i$ values and those from
\citet{abt02}.  The diamonds represent the measurements derived from 
spectra from the Elodie archive while the asterisks represent the corrected
measurements (see text) derived from spectra from the NOAO Indo-U.S.\ 
Library of Coud\'e Feed Stellar Spectra. The dotted line is the result
of a linear least-squares fit.  The most discrepant star in this
figure is HD~172958.}
\label{fig1}
\end{figure}

\placefigure{fig1}     

The effective temperature and gravity were derived by fitting the H$\gamma$
profile (see details in \citealt{hua06b}).  The results and the associated
numerical fitting errors are listed in columns (2) to (5) of Table~1.  As
pointed out by \citet{hua06b}, the derived $\log g$ values represent
an average of gravity over the visible hemisphere of these rotating stars.
They may not be good indicators of stellar evolutionary status, especially
for rapid rotators that have much lower gravity in the equatorial area caused
by the strong centrifugal force.  Following the method described in \citet{hua06b},  
we made a statistical correction to estimate the polar gravity of each star from its
derived $V \sin i$, $T_{\rm eff}$, and $\log g$, and the resulting polar gravity
is listed in column (8) of Table~1.  
Our estimates of $\log g_{\rm polar}$ are consistent with the available 
observations.  For example, one of our targets is Regulus (HD~87901) that was recently 
resolved by the CHARA Array optical long baseline interferometer \citep{mca05}.  
Models of the spectroscopy and interferometry of this rotationally
deformed star lead directly to a polar gravity of $\log g_{\rm polar}=3.98$, which
compares well with the statistical estimate here of $\log g_{\rm polar}=3.95$.
Furthermore, we used our derived $\log g_{\rm polar}$ values with 
masses estimated from Figure~3 to derive radii, luminosities, bolometric 
corrections, and absolute magnitudes.  We combined these with the observed magnitudes 
to find distance estimates, and a comparison of the derived distances 
with those from {\it Hipparcos} \citep{van07} shows good consistency.
We note for completeness that in a sample of
ten stars in common, \citet{fit05} find temperatures that are $\approx 4\%$ larger
and gravities that are $\approx 0.1$ dex greater than our values.  While
these differences between results from spectral flux and H$\gamma$ fitting
are interesting, they are insignificant for our purpose of comparing the
parameters of the field and cluster B-stars in a consistent manner.


\section{A Comparison of Field and Cluster B Stars}       

The recent studies \citep{abt02,str05,hua06a,wol07} that found that field
B-stars appear to rotate slower than cluster B-stars were mainly based on
the field sample from \citet{abt02} that includes roughly 1100 bright field
B-stars selected from the Bright Star Catalogue \citep{hof82}.
We note that both this and our own smaller sample of B-stars are not
volume-limited but tend to select from the intrinsically brighter members 
of the population.  Furthermore, both samples include some members of nearby
OB associations and moving groups, which are not strictly ``field'' objects. 
Nevertheless, these field samples are similar enough in their sampling
of the spectral types, luminosities, and true field star content that we 
can use both to compare with the cluster star rotational properties. 
The Abt et al.\ field sample (ALG02) contains a total of 902 B-stars of classes III-V,
excluding all SB2s, which we use in our statistical analysis below. 
Our field sample consists of only 108 B-stars, so one might question whether
its content and size are sufficient to represent a field star population
similar to that of ALG02.  The spectral sub-type distribution of our field sample 
and of the ALG02 sample are very similar (see Table~2).  Furthermore, 
we show in Figure~2 that the cumulative distribution functions of projected 
rotational velocities $V \sin i$ appear to be the same.   The mean $V \sin i$ 
of our field sample is $114\pm 9$ km~s$^{-1}$ while the mean $V \sin i$ of the 
ALG02 sample is $116\pm 3$ km~s$^{-1}$.  A Kolmogorov-Smirnov (KS) 
test shows that these two samples have a probability of 0.72 
to be drawn from the same parent sample.  Thus, we conclude that our 
limited sample makes a fair representation of the larger field sample of ALG02  
and of the rotational properties associated with this group of stars. 

\begin{figure}
\epsscale{1.}
\plotone{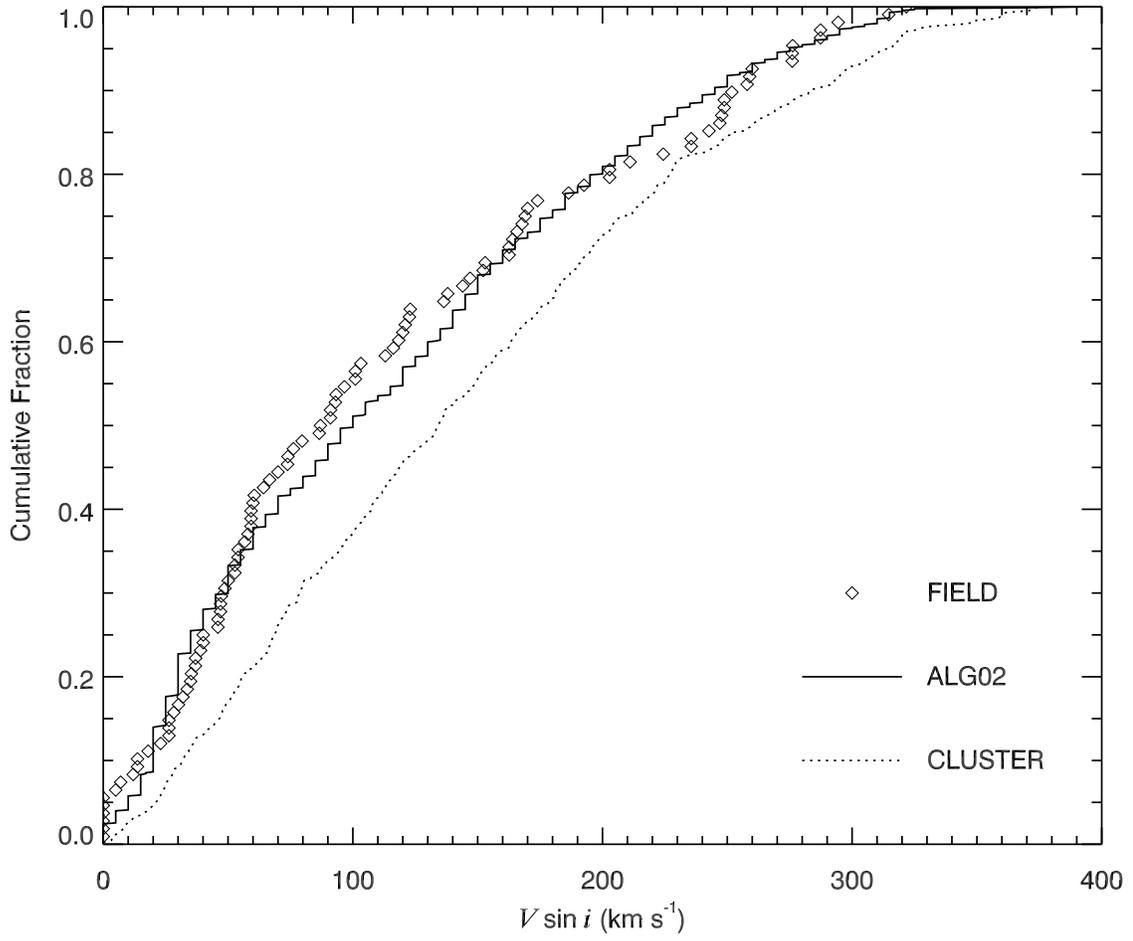}
\caption{The cumulative distribution function of projected 
rotational velocity for several different samples.}
\label{fig2}
\end{figure}

\placetable{tab2}      
\placefigure{fig2}     

The cluster B-star sample used for comparison is extracted from our 
previous survey of B-stars in 19 open clusters \citep{hua06a,hua06b}.  After removing 
all O-stars and SB2s, 432 cluster B-stars remain in this sample, which
covers a range of age from 6 to 72 Myr (the average is 12.5 Myr). 
The mean $V \sin i$ of the cluster sample is $146\pm 4$ km~s$^{-1}$,
which is definitely higher than the corresponding value of the
field sample.  The cumulative curve for the cluster sample
is also significantly different
from that of the field sample (Fig.~2).  The KS probability
that our field and cluster samples are drawn from the same parent sample
is as low as 0.001.  

The distributions of the field and the cluster B-star samples in the
$\log T_{\rm eff} - \log g_{\rm polar}$ plane are plotted in Figure~3.  
We see that along each evolutionary track the field stars are more 
evenly distributed in $\log g_{\rm polar}$ than is the case for 
the cluster stars, which mainly have higher $\log g_{\rm polar}$ (near the ZAMS).
This indicates that the field B-star sample contains a larger fraction of older stars 
(i.e., with lower $\log g_{\rm polar}$) than found in the cluster B-star 
sample\footnote{Some of the low $\log g_{\rm polar}$ stars among the 
young cluster sample may be pre-main sequence stars \citep{hua06b}.}.
If the stars in the field sample spin down with time in a similar way as 
those in the cluster sample \citep{hua06b}, then it is not surprising that 
the field sample with more older B-stars appears to be rotating slower than 
the cluster sample.  Note that the cluster sample contains relatively more massive 
stars compared to the field star sample because the cluster targets were 
typically selected from the brighter, more massive cluster members.  

\begin{figure}
\epsscale{1.}
\plotone{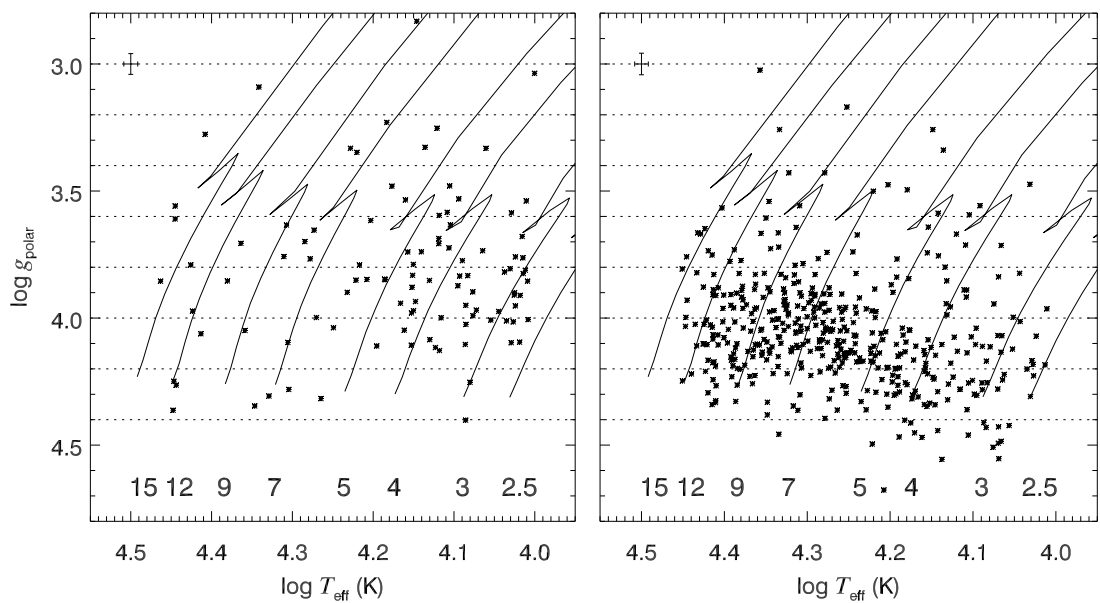}
\caption{The distribution of the sample B-stars in the $\log T_{\rm eff}-
\log g_{\rm polar}$ plane.  The left panel shows the distribution for the field
sample, and the right panel shows the same for the cluster sample.  The
average errors in $\log T_{\rm eff}$ and $\log g_{\rm polar}$ are
plotted in the top left corner of each panel.  The solid lines
are the evolutionary tracks for non-rotating stellar models 
\citep{sch92} marked by the initial mass ($M_\odot$) at the bottom.
}
\label{fig3}
\end{figure}

\placefigure{fig3}     

Is the larger fraction of older B-stars in the field sample 
the dominant cause of its apparent slow rotation 
or do some additional factors, such as the initial conditions and environment, 
need to be considered?
In order to investigate this, we plot in Figure~4 the $V \sin i$ distributions
of both the field and cluster samples against $\log g_{\rm polar}$. 
Figure 4 also illustrates the mean $V \sin i$ of stars in
each bin of 0.2 dex in $\log g_{\rm polar}$ ({\it solid line}) and the
associated standard deviation of the mean ({\it shaded area}).  
The advantage of using Figure~4 is that the evolutionary
spin down effect is dramatically revealed as we compare the stellar rotation
of the two samples in each $\log g_{\rm polar}$ bin.  
The overall decrease in  mean $V \sin i$ with lower 
$\log g_{\rm polar}$ shows clearly 
that the spin down process exists in both samples.  
By comparing the mean $V \sin i$ of corresponding bins, 
we found that it is difficult to draw a firm conclusion 
about which sample rotates faster.  At each evolutionary
stage (indicated by $\log g_{\rm polar}$), the B-stars in
these two samples appear to rotate equally fast.  Thus, the overall
slowness of rotation in the field sample is mainly due to
the larger percentage of its content occupying the bins of lower 
$\log g_{\rm polar}$.

\begin{figure}
\epsscale{0.75}
\plotone{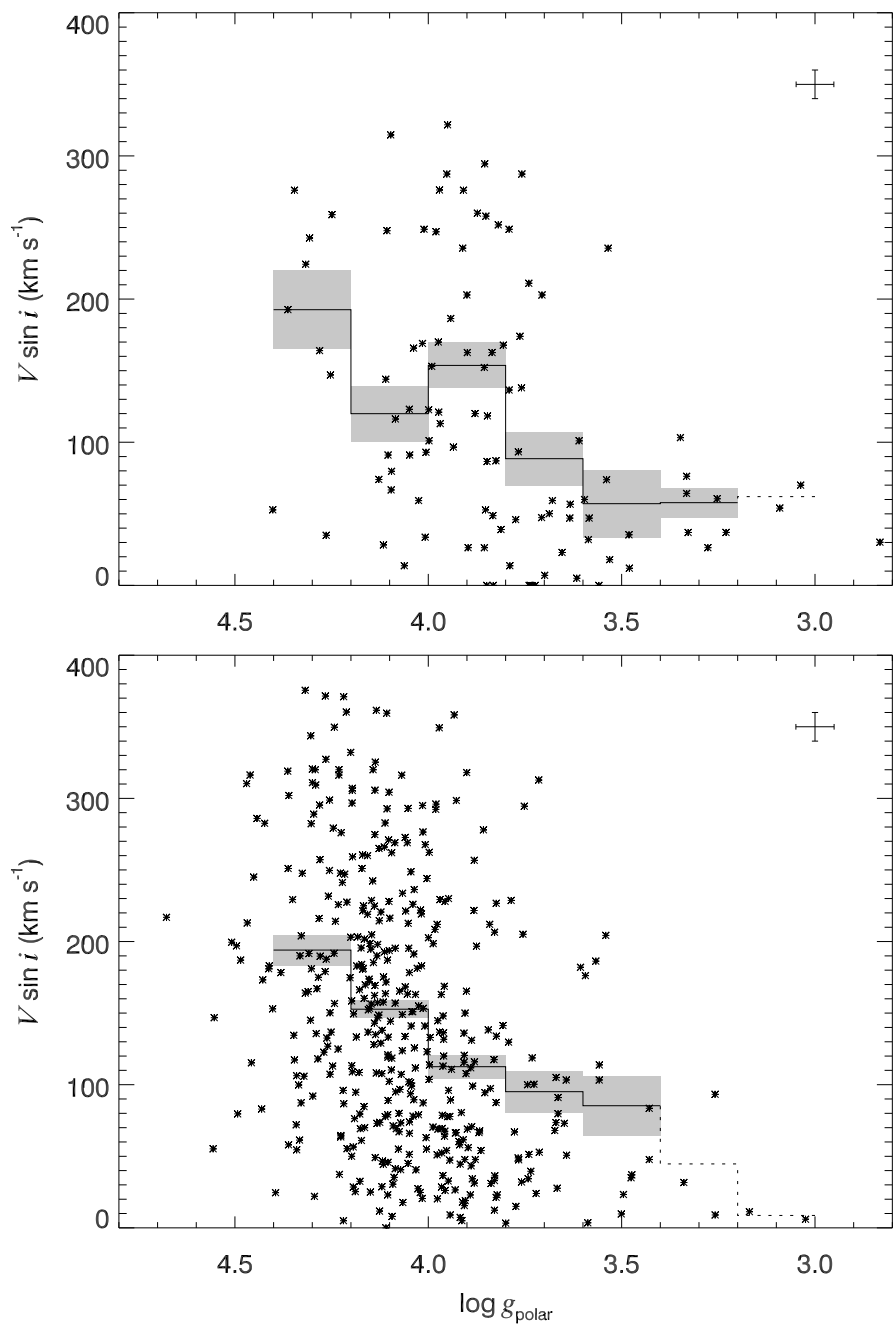}
\caption{The distribution of the field ({\it top}) and cluster ({\it bottom})
sample B-stars in the $\log g_{\rm polar}-V \sin i$ plane. The average
errors in $\log g_{\rm polar}$ and $V \sin i$ are plotted in the top
right corner of each panel.  The solid line shows the mean $V \sin i$ of each
0.2 dex bin of $\log g_{\rm polar}$ that contains six or more measurements
while the dotted line shows the same for the rest of bins. The
shaded areas indicate the associated error of the mean in each bin.
}
\label{fig4}
\end{figure}

\placefigure{fig4}     

Note that we are interpreting the line broadening solely in terms
of rotation, but \citet{rya02} and \citet{duf06} find that 
macroturbulent broadening is also important among the luminous
supergiants (where it may amount to velocities of 20 -- 60 km~s$^{-1}$). 
We have only seven stars in the sample with $\log g_{\rm polar} < 3.0$, 
and these have measured $V \sin i$ velocities of 31 -- 59 km~s$^{-1}$, 
i.e., comparable to the expected macroturbulent velocities. 
Thus, we regard the $V \sin i$ values of the stars with low 
$\log g_{\rm polar}$ as upper limits, and the trend of declining 
rotation velocity with lower $\log g_{\rm polar}$ may actually 
be steeper than indicated in the low $\log g_{\rm polar}$ part 
of Figure~4. 

One possible concern about the comparison made above is that many late B-stars
in our cluster sample are found to have non-solar helium abundances \citep{hua06b}.
Since the hydrogen abundance will be lower in helium enriched atmospheres, 
the change in atmospheric opacity may cause a change in the appearance
of the H$\gamma$ profile that could lead to erroneous derived values of 
$T_{\rm eff}$ and $\log g$.   We checked this possibility by measuring 
the H$\gamma$ profile in synthetic model spectra for He-peculiar 
stars\footnote{http://star.arm.ac.uk/\%7Ecsj/models/Grid.html}
calculated by C.\ S.\ Jeffrey using the {\it Sterne/Spectrum} LTE codes
\citep*{jef01}.  Our results are shown in Table~3 that lists the 
fraction of H and He atoms by number and our derived $T_{\rm eff}$ and 
$\log g$ for three temperature cases.  The three rows in the table give
the results for sub-solar He, solar He, and enhanced He, respectively. 
Ideally, we should recover exactly the assumed model $T_{\rm eff}$ and $\log g$
for the solar He case, but our scheme arrives at temperatures 
that are somewhat low (especially at higher $T_{\rm eff}$; for the
expected values of 16000K/4.0, we obtain derived values of 15200K/3.95).
We suspect that this systematic difference reflects differences between the
LTE codes {\it Sterne/Spectrum} and the LTE codes {\it ATLAS9/SYNSPEC} 
that we used to develop the H$\gamma$ calibration.  While these 
differences are significant, they are not important for our analysis here
where we are making a differential comparison between the cluster and field samples
using the same method to obtain the stellar parameters.  What is important
are the relative changes as the He abundance increases.  We see that 
He enrichment results in a deeper H$\gamma$ profile that is interpreted 
in our scheme mainly as a decrease in the resulting temperature while 
changes in the derived gravity are small.  Furthermore, we show in 
Figure~5 that we find no evidence of a correlation between He abundance
and $V\sin i$ among the late B stars ($T_{\rm eff}< 20000K$) in our cluster sample.
Thus, any corrections to the gravity that might be applied to the He-peculiar 
star subset would be too small to change the rotational trends seen in Figure~4.  

\begin{figure}
\epsscale{0.8}
\plotone{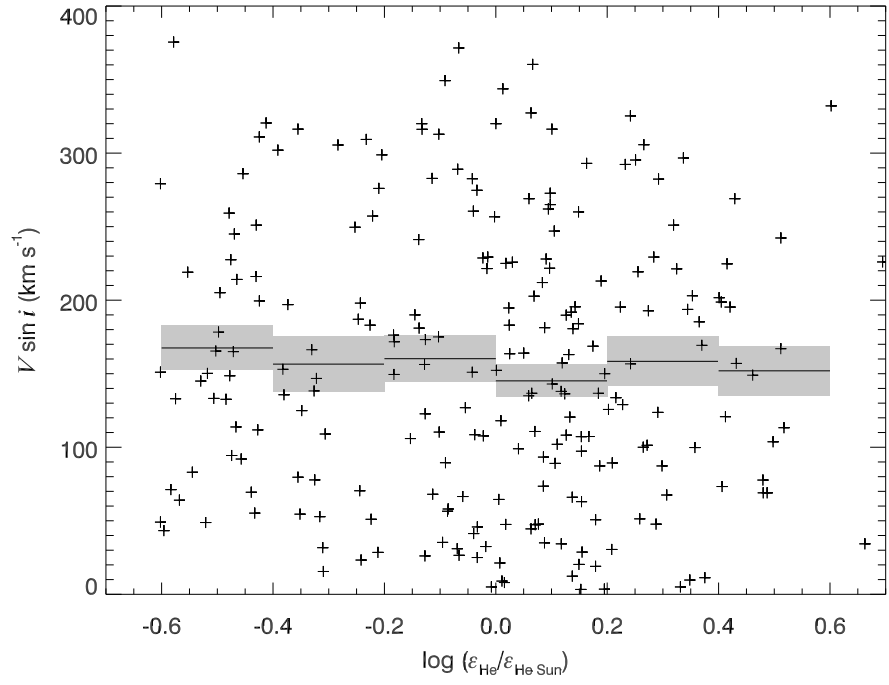}
\caption{The distribution of the late B-stars ($T_{\rm eff}< 20000$ K) in our
cluster sample in the $\log (\epsilon_{\rm He}/\epsilon_{\rm He\odot})
-V \sin i$ plane. The thick lines indicate the mean $V \sin i$ value
of each 0.2 dex bin of $\log (\epsilon_{\rm He}/\epsilon_{\rm He\odot})$.
The shadowed areas show the error of the mean in each bin.
}
\label{fig5}
\end{figure}

\placetable{tab3}      
\placefigure{fig5}     


\section{Discussion and Conclusions}                             

Our findings in previous section seem to support the first
explanation mentioned in \S1, i.e., 
field B-stars rotate slower statistically because they represent 
an older population than cluster B-stars.
The projected rotational velocities with each $\log g_{\rm polar}$
bin (corresponding to evolutionary state) appear to be very 
similar in the field and cluster samples, which suggests that 
any differences in environmental factors at birth between the
field and cluster samples has little influence on their present
rotational properties.  This conclusion differs from that of
\citet{str05} and \citet{wol07} who argue that the stellar
number density at formation affects the rotational velocity
distribution.

Our field sample contains 108 B-stars.  The relatively small
sample size precludes an analysis of subsets based on binned mass
ranges.  This raises a question: can we use the whole field sample 
of B-stars, which span a large range of mass and main sequence (MS) 
lifetime, to compare with the cluster sample, and still draw 
meaningful conclusions?   The answer is yes, since our method of 
comparison is based on the estimated polar gravity $\log g_{\rm polar}$ 
of individual B-stars.  For all subtypes of MS B-stars, 
the surface $\log g$ falls in a range between 4.2 -- 4.3 
(for the zero-age main sequence, ZAMS) to 3.4 -- 3.6 (for the 
terminal-age main sequence, TAMS), as shown in Figure 3.  
The evolutionary spin down of a MS B-star is mainly due to the
evolutionary increase of its moment of inertia (and stellar
radius) and/or stellar wind mass loss.  However, compared to the 
more massive O-stars, the stellar winds of MS B-stars are generally
weak, so wind mass loss plays a minor role in spin down.  Thus, the 
evolutionary changes in stellar properties, such as stellar radius and
moment of inertia, will be the major cause of evolutionary
spin down.  These properties are directly related to surface
$\log g$ of the star (or more accurately, $\log g_{\rm polar}$ for a
rotating star).  In this sense, consideration of B-stars binned in 
groups of similar $\log g_{\rm polar}$ is a reasonable means to 
search for evidence of changes in the mean rotational properties 
with advancing evolutionary state.

\citet{str05} relied on the Str\"omgren $\beta$ and $c_0$ indices
to select their objects in both the field and cluster samples.  
We note, however, that estimates of surface gravity derived from fitting
the H$\gamma$ line profile are generally more reliable than those 
based upon $\beta$ index, which has a larger intrinsic error
(since $\beta$ measures the difference in magnitude 
between a narrow band and a wide band centered 
at H$\beta$).  Thus, even though the
cluster and field samples were selected from the same area
in the $\beta - c_0$ plane, they may still contain populations in
different evolutionary states (i.e., over a greater range in gravity).  
The cluster sample from \citet{str05} is known
to be young because it consists of member stars from young open clusters
(h and $\chi$ Persei) while the field sample may include a lot
of older stars because its content relies on the $\beta - c_0$
selection criterion.  
\citet{sma95} calculated a grid of synthetic $\beta$ indices 
applicable to B-stars (given in Table 7 of their paper).  
The differences in $\beta$ index between ZAMS ($\log g = 4.0$)
and TAMS ($\log g = 3.5$) B-stars are only about 0.04 -- 0.05 mag.
Since the Str\"{o}mgren data collected by \citet{str05} 
for the field B-stars came from diverse sources and have errors 
of 0.01 - 0.02 mag, it is not easy to distinguish between 
the evolved and unevolved stars based upon the $\beta$ index alone. 
Thus, despite their best efforts to compare the rotational velocities
of comparably evolved stars in $h$ and $\chi$~Per and the field,
\citet{str05} probably included a significant fraction of more 
evolved stars in the field sample.
Our field sample has 21 stars in common with
the low mass group (group 1) of the field sample from 
\citet{str05}, the group with the largest difference in the 
$V \sin i$ cumulative distribution from their cluster sample.  
Among these 21 stars, 14 have $\log g_{\rm polar} < 4.0$.  
Figure~3 shows that the majority of cluster B-stars with
mass less than 5 $M_\odot$ has $\log g_{\rm polar} > 4.0$.
If we assume that the rest of field B-stars in their group 1
are similar to these 21 stars, the slower rotation in
group 1 of their field sample can be naturally explained by
its older population, instead of the initial conditions (a low
density environment of the star forming region) as suggested
in their paper. 

\citet{wol07} investigated stellar samples from both
low density and high density stellar environments.  In their analysis,
they first inspected the evolutionary effect (spin-down) 
on stellar rotation, and concluded that the evolutionary
effect is too small to account for the difference in
stellar rotation that exists between the low and high density 
samples.  However, the evolutionary status of individual
stars in their samples is based on the estimated age
of the parent association or cluster only.  This approach to
evolutionary change is less specific than our estimate based
upon the polar gravity of each star, since the individual
cluster samples may contain quite different proportions of
evolved to unevolved stars. Thus, it is possible that 
the samples considered by \citet{wol07} contain stars 
that occupy a wider range of evolutionary state than assumed.
Consequently, their comparison between the low-density
and high-density cumulative probability curves that 
are based on the whole sample may be influenced more by the 
evolutionary effect on stellar rotation than the authors realized.

In summary, our spectroscopic investigation of the stellar rotation
of 108 field B-stars suggests that the field B-stars contain a 
larger fraction of more evolved stars than found among our sample
of young cluster stars (with an average age of 12.5 Myr)
and that makes the field stars appear to rotate
slower as whole.  This is not a surprising result, since 
most of the bright field stars belong to the local Gould's Belt structure
that has an expansion age of 30 to 60~Myr \citep*{tor00}.
At this point, we do not see any significant differences
between the rotational distributions of the field and young cluster
B-stars when considered as a function of evolutionary state.
We applied identical spectroscopic
methods to both the field and cluster samples, 
and this should minimize any method-related errors in the 
comparison of rotational properties.  We used the estimated
$\log g_{\rm polar}$ as an indicator of evolutionary
status for each individual star, a
necessary precaution for rapidly rotating stars and for the purpose
of our paper.  Our field B-star sample is
still small.  In the near future, we plan to obtain more 
spectra of a much larger field B-star sample to improve 
the statistical basis of our conclusion and to 
investigate the subgroups in confined stellar mass ranges.


\acknowledgments

The spectral data used in this paper are from the NOAO Indo-U.S.\ Library
of Coud\'e Feed Stellar Spectra and the ELODIE archive. 
This material is based upon work supported by the National Science Foundation 
under Grant No.~AST-0606861.  The authors are also very grateful for partial 
support from NSF grant No.~AST-0507219 to Dr. Judith G. Cohen.




\clearpage


\begin{deluxetable}{lcccccccl}
\tablewidth{0pc}
\tablecaption{Derived Stellar Parameters
\label{tab1}}
\tablehead{
\colhead{ } &
\colhead{$T_{\rm eff}$ } &
\colhead{$\Delta T_{\rm eff}$ } &
\colhead{ } &
\colhead{ } &
\colhead{$V \sin i$ } &
\colhead{$\Delta V \sin i$ } &
\colhead{ } &
\colhead{Spec. } \\
\colhead{HD } &
\colhead{(K) } &
\colhead{(K) } &
\colhead{$\log g$ } &
\colhead{$\Delta\log g$ } &
\colhead{(km~s$^{-1})$ } &
\colhead{(km~s$^{-1})$ } &
\colhead{$\log g_{\rm polar}$ } &
\colhead{Class.}}
\startdata
   886  &   19255  & 294  & 3.696  & 0.034  &  \phn\phn7\tablenotemark{a}  &   \phn5  & 3.699  &  B2 IV \\
  3360  &   18755  & 350  & 3.642  & 0.043  &   \phn23\tablenotemark{a}  &   \phn4  & 3.654  &  B2 IV \\
 10362  &   13211  & 133  & 3.144  & 0.024  &   \phn61  &  11  & 3.253  &  B7 II \\
 12303  &   11491  & \phn81  & 3.195  & 0.023  &   \phn77  &  12  & 3.332  &  B8 III \\
 17081  &   12769  & \phn89  & 3.689  & 0.023  &  \phn\phn5  &  20  & 3.724  &  B7 IV \\
 18296  &   11602  & 146  & 3.702  & 0.047  &  \phn\phn0  &  16  & 3.702  &  B9p \\
 24398  &   21950  & 504  & 3.061  & 0.055  &   \phn54\tablenotemark{a}  &   \phn5  & 3.091  &  B1 Iab \\
 24760  &   26517  & 648  & 3.923  & 0.058  &  121\tablenotemark{a}  &   \phn9  & 3.973  &  B0.5 V \\
 25940  &   17746  & 552  & 3.898  & 0.060  &  166  &  10  & 4.038  &  B3 Ve \\
 27295  &   11334  & 113  & 3.972  & 0.036  &   \phn33  &  15  & 4.008  &  B9 IV \\
 33904  &   12291  & 135  & 3.715  & 0.040  &   \phn46  &  14  & 3.774  &  B9 IV \\
 34816  &   25892  & 714  & 4.053  & 0.076  &   \phn13  &  14  & 4.062  &  B0.5 IV \\
 35468  &   20286  & 411  & 3.613  & 0.051  &   \phn47\tablenotemark{a}  &   \phn5  & 3.634  &  B2 III \\
 35497  &   13129  & \phn98  & 3.537  & 0.023  &   \phn60\tablenotemark{a}  &   \phn5  & 3.596  &  B7 III \\
 38899  &   10272  & \phn40  & 3.781  & 0.018  &   \phn39\tablenotemark{a}  &   \phn4  & 3.812  &  B9 IV \\
 40111  &   27866  & 535  & 3.559  & 0.059  &  101\tablenotemark{a}  &   \phn2  & 3.610  &  B0.5 II \\
 41692  &   13669  & 144  & 3.260  & 0.020  &   \phn37  &  12  & 3.328  &  B5 IV \\
 43247  &   10391  & \phn72  & 2.573  & 0.025  &   \phn40  &  13  & 2.700  &  B9 II-III \\
 51309  &   16898  & 406  & 2.657  & 0.047  &   \phn59  &  14  & 2.766  &  B3Ib/II \\
 58343  &   15025  & 317  & 3.428  & 0.045  &   \phn35  &  10  & 3.481  &  B2 Vne \\
 74280  &   18630  & 411  & 3.933  & 0.050  &  101\tablenotemark{a}  &   \phn5  & 3.998  &  B3 V \\
 75333  &   12105  & 121  & 3.775  & 0.036  &   \phn49  &  16  & 3.833  &  B9mnp \\
 79158  &   12718  & 228  & 3.554  & 0.056  &   \phn57  &  12  & 3.633  &  B8mnp III\\
 79469  &   10190  & \phn39  & 3.920  & 0.022  &   \phn93\tablenotemark{a}  &   \phn7  & 4.006  &  B9.5 V \\
 87344  &   10689  & \phn64  & 3.526  & 0.026  &   \phn32  &   \phn9  & 3.586  &  B8 V \\
 87901  &   12174  & \phn63  & 3.574  & 0.018  &  322  &  11  & 3.950  &  B7 V \\
100889  &   10422  & \phn38  & 3.649  & 0.018  &  235  &  10  & 3.911  &  B9.5 Vn \\
116658  &   28032  & 868  & 4.301  & 0.109  &  192  &  14  & 4.363  &  B1 III-IV \\
120315  &   15689  & 128  & 4.004  & 0.022  &  144\tablenotemark{a}  &   \phn5  & 4.110  &  B3 V \\
129956  &   10333  & \phn51  & 3.731  & 0.023  &   \phn87\tablenotemark{a}  &   \phn7  & 3.825  &  B9.5 V \\
135742  &   12450  & 226  & 3.565  & 0.065  &  260  &  26  & 3.873  &  B8 V \\
145502  &   20157  & 295  & 4.194  & 0.039  &  164\tablenotemark{a}  &   \phn8  & 4.281  &  B3 V/B2 IV\\
147394  &   14166  & 149  & 3.806  & 0.026  &   \phn\phn0  &  15  & 3.806  &  B5 IV \\
149630  &   10600  & \phn34  & 3.598  & 0.017  &  276\tablenotemark{a}  &  15  & 3.909  &  B9 V \\
150100  &   10441  & \phn42  & 4.015  & 0.017  &   \phn79  &  12  & 4.095  &  B9.5 Vn \\
150117  &   10594  & \phn37  & 3.670  & 0.017  &  203  &  10  & 3.900  &  B9 V \\
152614  &   11812  & \phn41  & 3.865  & 0.013  &  113\tablenotemark{a}  &   \phn4  & 3.969  &  B8 V \\
154445  &   22831  & 363  & 3.985  & 0.034  &  123\tablenotemark{a}  &   \phn5  & 4.049  &  B1 V \\
155763  &   12833  & \phn86  & 3.543  & 0.020  &   \phn47\tablenotemark{a}  &   \phn4  & 3.584  &  B6 III \\
157741  &   10569  & \phn43  & 3.639  & 0.020  &  287  &  13  & 3.952  &  B9 V \\
158148  &   14210  & \phn99  & 3.733  & 0.017  &  247\tablenotemark{a}  &   \phn6  & 3.980  &  B5 V \\
160762  &   15961  & 155  & 3.613  & 0.025  &  \phn\phn5\tablenotemark{a}  &   \phn2  & 3.616  &  B3 IV \\
161056  &   20441  & 327  & 3.433  & 0.039  &  287  &   \phn8  & 3.758  &  B1.5 V \\
164284  &   22211  & 573  & 4.207  & 0.055  &  276\tablenotemark{a}  &   \phn7  & 4.346  &  B2 Ve \\
164353  &   15488  & 334  & 2.638  & 0.036  &   \phn46\tablenotemark{a}  &   \phn9  & 2.694  &  B5 Ib \\
166014  &   10345  & \phn28  & 3.511  & 0.020  &  174\tablenotemark{a}  &  12  & 3.763  &  B9.5 V \\
168199  &   14660  & 104  & 3.762  & 0.019  &  186  &   \phn8  & 3.942  &  B5 V \\
168270  &   10245  & \phn34  & 3.419  & 0.018  &   \phn74  &  10  & 3.539  &  B9 V\\
169578  &   10901  & \phn36  & 3.498  & 0.014  &  252  &   \phn9  & 3.819  &  B9 V \\
171301  &   12170  & \phn82  & 3.969  & 0.025  &   \phn59  &  13  & 4.025  &  B8 IV \\
171406  &   14216  & 115  & 3.881  & 0.022  &  248  &  10  & 4.107  &  B4 Ve \\
172958  &   10727  & \phn69  & 3.577  & 0.030  &  167  &  12  & 3.806  &  B8 V \\
173087  &   14504  & 111  & 3.970  & 0.025  &   \phn91  &  10  & 4.048  &  B5 V \\
173936  &   13489  & \phn88  & 3.989  & 0.015  &  116  &   \phn8  & 4.085  &  B6 V \\
174959  &   13499  & \phn80  & 3.795  & 0.012  &   \phn52  &  11  & 3.852  &  B6 IV \\
175156  &   14001  & \phn77  & 2.753  & 0.013  &   \phn31  &  14  & 2.832  &  B3 II \\
175426  &   16137  & 197  & 3.764  & 0.032  &   \phn86  &  10  & 3.848  &  B2.5 V \\
175640  &   11932  & 141  & 3.861  & 0.046  &   \phn27  &  13  & 3.897  &  B9 III \\
176318  &   13058  & \phn67  & 3.888  & 0.015  &  122  &   \phn8  & 3.999  &  B7 IV \\
176437  &   10005  & \phn48  & 2.909  & 0.026  &   \phn70\tablenotemark{a}  &   \phn9  & 3.037  &  B9 III \\
176582  &   15338  & 150  & 3.727  & 0.024  &  119  &  13  & 3.847  &  B5 IV \\
176819  &   20209  & 356  & 4.056  & 0.039  &   \phn67  &  10  & 4.096  &  B2 IV-V \\
177756  &   11084  & \phn41  & 3.822  & 0.016  &  170\tablenotemark{a}  &   \phn5  & 3.974  &  B9 Vn \\
177817  &   12387  & \phn55  & 3.642  & 0.019  &  162  &  12  & 3.835  &  B7 V \\
178125  &   13120  & 100  & 4.078  & 0.017  &   \phn74\tablenotemark{a}  &   \phn7  & 4.128  &  B8 III \\
178329  &   15317  & 208  & 3.827  & 0.033  &   \phn\phn0  &  19  & 3.827  &  B3 V \\
179588  &   12177  & 101  & 4.366  & 0.033  &   \phn52  &  14  & 4.402  &  B9 IV \\
179761  &   12746  & 103  & 3.469  & 0.027  &   \phn12\tablenotemark{a}  &   \phn6  & 3.480  &  B8 II-III \\
180163  &   15250  & 164  & 3.196  & 0.026  &   \phn37\tablenotemark{a}  &   \phn7  & 3.230  &  B2.5 IV \\
180968  &   27974  & 731  & 4.141  & 0.107  &  259  &   \phn7  & 4.249  &  B0.5 IV \\
182568  &   16479  & 219  & 3.653  & 0.035  &  137  &   \phn8  & 3.791  &  B3 IV \\
183144  &   14361  & 126  & 3.484  & 0.028  &  211  &   \phn8  & 3.740  &  B4 III \\
184915  &   26654  & 747  & 3.592  & 0.072  &  249  &   \phn7  & 3.791  &  B0.5 III \\
184930  &   13148  & \phn89  & 3.621  & 0.016  &   \phn50  &   \phn9  & 3.687  &  B5 III \\
185423  &   16603  & 328  & 3.209  & 0.049  &  103  &  14  & 3.348  &  B3 III \\
185859  &   25577  & 625  & 3.264  & 0.041  &   \phn27  &  23  & 3.277  &  B0.5 Iae \\
187811  &   21331  & 640  & 4.173  & 0.062  &  242  &  10  & 4.307  &  B2.5 Ve \\
187961  &   16646  & 441  & 3.554  & 0.063  &  258  &  10  & 3.851  &  B7 V \\
188260  &   10363  & \phn50  & 3.592  & 0.025  &   \phn59  &   \phn8  & 3.679  &  B9.5 III \\
189944  &   14134  & 175  & 3.758  & 0.035  &   \phn12  &  15  & 3.789  &  B4 V \\
191243  &   14368  & 285  & 2.580  & 0.049  &   \phn55  &  13  & 2.703  &  B5 Ib \\
191639  &   29047  &1343  & 3.777  & 0.157  &  152  &  15  & 3.855  &  B1 V \\
192276  &   13272  & 155  & 4.088  & 0.031  &   \phn29  &  12  & 4.116  &  B7 V \\
192685  &   17062  & 242  & 3.746  & 0.033  &  162  &  11  & 3.899  &  B3 V \\
193432  &   10208  & \phn53  & 3.814  & 0.028  &   \phn27  &  18  & 3.855  &  B9 IV \\
195810  &   13146  & 121  & 3.646  & 0.025  &   \phn47  &  10  & 3.707  &  B6 III \\
196504  &   10693  & \phn59  & 3.781  & 0.026  &  315  &  13  & 4.097  &  B9 V \\
196740  &   14129  & 154  & 3.673  & 0.030  &  276  &   \phn7  & 3.971  &  B5 IV \\
196867  &   10568  & \phn44  & 3.572  & 0.017  &  138\tablenotemark{a}  &   \phn5  & 3.759  &  B9 IV \\
198183  &   14187  & 137  & 3.765  & 0.027  &  120\tablenotemark{a}  &  10  & 3.879  &  B5 Ve \\
205021  &   27784  & 768  & 4.261  & 0.064  &   \phn35\tablenotemark{a}  &   \phn5  & 4.264  &  B2 IIIe\\
205139  &   27860  & 540  & 3.556  & 0.057  &  \phn\phn0  &  24  & 3.556  &  B1 II \\
205637  &   23102  & 951  & 3.515  & 0.073  &  203  &   \phn8  & 3.706  &  B3 Vp \\
206165  &   19887  & 394  & 2.730  & 0.046  &   \phn58\tablenotemark{a}  &  12  & 2.782  &  B2 Ib \\
207330  &   16908  & 231  & 3.243  & 0.032  &   \phn64  &  18  & 3.332  &  B3 III \\
207516  &   12187  & \phn80  & 4.020  & 0.024  &   \phn91  &  10  & 4.104  &  B8 V \\
208501  &   17369  & 194  & 2.492  & 0.043  &   \phn40  &  25  & 2.571  &  B8 Ib \\
209409  &   18389  & 524  & 4.178  & 0.065  &  224  &   \phn8  & 4.317  &  B7 IVe \\
209419  &   13815  & 121  & 3.708  & 0.025  &  \phn\phn0  &  12  & 3.708  &  B5 III \\
209819  &   12026  & \phn45  & 4.161  & 0.013  &  147  &   \phn8  & 4.253  &  B8 V \\
212571  &   24011  & 713  & 3.593  & 0.071  &  294  &   \phn8  & 3.854  &  B1 Ve \\
212978  &   18966  & 248  & 3.682  & 0.029  &   \phn93  &   \phn9  & 3.767  &  B2 V \\
214923  &   11927  & \phn89  & 3.858  & 0.030  &  153\tablenotemark{a}  &   \phn3  & 3.991  &  B8 V \\
217675  &   14458  & 210  & 3.195  & 0.040  &  235  &  11  & 3.535  &  B6 IIIpe \\
220575  &   12419  & 125  & 3.514  & 0.034  &   \phn18\tablenotemark{a}  &   \phn5  & 3.531  &  B8 III \\
222439  &   10632  & \phn41  & 3.875  & 0.019  &  169\tablenotemark{a}  &   \phn4  & 4.015  &  B9 IVn \\
224926  &   14047  & 118  & 3.842  & 0.023  &   \phn97  &  26  & 3.935  &  B7 III-IV \\
225132  &   10839  & \phn48  & 3.767  & 0.014  &  249  &  10  & 4.011  &  B9 IVn \\

\enddata
\tablenotetext{a}{Derived $V \sin i$ using spectra from the Elodie library.} 
\end{deluxetable}


\begin{deluxetable}{lcccc}
\tablewidth{0pc}
\tablecaption{Field Sample Spectral Distribution
\label{tab2}}
\tablehead{
\colhead{Sample } &
\colhead{B0-2 } &
\colhead{B3-5 } &
\colhead{B6-8 } &
\colhead{B9-9.5 }}
\startdata
ALG02     &  23.6\%  &  20.8\%  &   28.5\%  &    27.1\%   \\
This work &  23.1\%  &  25.9\%  &   26.9\%  &    24.1\%   \\
\enddata
\end{deluxetable}


\begin{deluxetable}{cccccc}
\tablewidth{0pc}
\tablecaption{Tests of Derived $T_{\rm eff}$ and $\log g$ for He-peculiar
Model Spectra
\label{tab3}}
\tablehead{
\multispan{2}{Model Abundance} &
\multispan{4}{\hfil Tested Cases\hfil} \\
\\
\cline{1-2}
\cline{4-6}
\colhead{H} &
\colhead{He} &
\colhead{}&
\colhead{12(kK)/4.0} &
\colhead{14(kK)/4.0} &
\colhead{16(kK)/4.0} \\
\colhead{Fraction} &
\colhead{Fraction} &
\colhead{} &
\colhead{$\triangle T_{\rm eff}(\%)/\triangle\log g$ (dex)\tablenotemark{a}} &
\colhead{$\triangle T_{\rm eff}(\%)/\triangle\log g$ (dex)\tablenotemark{a}} &
\colhead{$\triangle T_{\rm eff}(\%)/\triangle\log g$ (dex)\tablenotemark{a}}}
\startdata
0.95 & 0.05 &&  +0.9/--0.01 &  +0.7/--0.05 &  +1.2/--0.03 \\
0.90 & 0.10 &&   0.0/  0.00 &   0.0/  0.00 &   0.0/  0.00 \\
0.70 & 0.30 && --5.3/  0.00 & --3.3/ +0.10 & --3.9/ +0.09 \\
\enddata
\tablenotetext{a}{The relative differences are calculated against the derived values
of the solar model (0.90 H and 0.10 He), which are 11900K/4.02, 13600K/4.00, and
15200K/3.95.}
\end{deluxetable}


\clearpage

\clearpage

\clearpage

\clearpage

\clearpage

\end{document}